\documentclass{workshop}
\begin{document}

\title{
	Cyclotron absorption lines in the era of Suzaku and NuSTAR
}

\author{
Gaurava K. Jaisawal and Sachindra Naik \\ 
\\[12pt]  
%
Astronomy and Astrophysics Division, Physical Research Laboratory, \\
Navrangapura, Ahmedabad - 380009, Gujarat, India \\
%
\textit{E-mail(GKJ): gaurava@prl.res.in} 
}

\abst{
Cyclotron resonance scattering features or cyclotron absorption lines are unique features observed in the hard X-ray spectra of accretion powered X-ray pulsars with magnetic field of the order of 10$^{12}$~G. Detection of these features enables us for the direct estimation of strength of the magnetic field close to the neutron star surface. Corresponding to magnetic field of $\sim$10$^{12}$~G, the fundamental cyclotron lines are expected in 10-100 keV energy range with harmonics expected at multiples of fundamental line energy. However, we detected first harmonic of cyclotron line at less than twice of the fundamental line energy ($\sim$1.7 times the fundamental line energy) in Be/X-ray binary pulsar Cep~X-4. With the broadband spectral capability of $Suzaku$ and $NuSTAR$ observatories, we have investigated cyclotron resonance scattering features in several X-ray pulsars to understand the shape of the lines, width, magnetic field mapping, anharmonicity in the line energies and luminosity-dependent properties of cyclotron lines. The results obtained from these works and new detection of cyclotron line in unknown/poorly studied sources are presented
in this paper.
}

\kword{stars: neutron -- pulsars : individual (4U 1909+07, Cep X-4, SMC X-2) -- accretion, magnetic field -- X-rays: stars
}

\maketitle
\thispagestyle{empty}

\section{Introduction}

Accretion powered X-ray pulsars are known to be highly magnetized rotating neutron 
stars. These sources are powered by the accretion of matter from the binary companion
either through Roche lobe overflow or capture of stellar wind from the massive companion. 
The broadband spectrum in 0.1-100 keV range showed the presence of several features
such as fluorescence emission lines, soft X-ray excess, broad absorption like features 
known as cyclotron resonance scattering features (CRSFs). CRSFs are generally 
detected in the hard X-ray spectrum of pulsars with magnetic field in the order 
of 10$^{12}$~G  (M{\'e}sz{\'a}ros 1992). These line-like features are formed due 
to the resonant scattering of photons with electrons near the neutron star surface. 
Depending on the magnetic field strength, the energy states of the electron are
quantized harmonically in Landau levels with an energy separation of 
E$_{cyc}$=11.6~B$_{12}\times(1+z){^{-1}}$ (keV), where B$_{12}$ is the magnetic 
field in the unit of 10$^{12}$~G and $z$ is the gravitational red-shift. Detection 
of CRSFs in the pulsar spectrum, thus, is a powerful tool for the direct estimation 
of magnetic field of the neutron star. After the launch of $Suzaku$ and $NuSTAR$, 
the number of cyclotron line sources has been rapidly increased and helped us in 
understanding the distribution of magnetic field lines around the poles of neutron 
stars. To date, detection of cyclotron lines are confirmed in 30 sources. 
However, there are about 8 sources in which these lines are tentatively detected 
(Table~\ref{cyc}). Apart from the fundamental line, harmonics of cyclotron lines are 
also seen in some cases. Many interesting aspects related to the line shape, 
anharmonicity in coupling factor, positive/negative dependence of line energy with 
luminosity have been studied in recent years. Among these, we present some 
recent results on cyclotron line obtained from the studies of X-ray pulsars 
such as 4U~1909+07, Cep~X-4 and SMC~X-2 by using $Suzaku$ and $NuSTAR$ 
observations in this article.

\begin{table}[!ht]
\centering\scriptsize
\caption{Up to date list of cyclotron line sources.}
\begin{tabular}{llll}
\hline
\hline
\\
   &Source 		          &CRSFs    	                            		 &Ref. \\
   &	               & (keV)	     	                &\\
\hline
\\
1 &Swift~J1626.6-5156      &10                                            	&DeCesar et al. (2013)\\
2 &XMMU J05$^*$                &10                                          &Manousakis et al. (2009) \\
3 &KS~1947+300             &12.5                                     &F{\"u}rst et al. (2014) \\
4 &4U 0115+634            &14, 24, 36                                 &Wheaton et al. (1979) \\ 
&                          &48, 62                                            &Ferrigno et al. (2011) \\
5  &IGR J17544-2619         &17                                             &Bhalerao et al. (2015)\\
6 &4U 1907+09              &19, 40                                         &Rivers et al. (2010) \\
7  &4U 1538-52              &22, 47                                           &Rodes-Roca et al. (2009) \\
8 &IGR J18179-1621         &22                                       &Li et al. (2012)   \\
9 &IGR J18027-2016         &23                                             &Lutovinov et al. (2017)  \\
10 &2S 1553-542             &23.5                                           &Tsygankov et al. (2016)\\
11 &Vela X-1                &25, 50                                         &La Barbera et al. (2003)  \\     
12 &V 0332+53               &27, 51, 74                                   &Nakajima et al. (2010) \\ 
13 &SMC~X-2                 &27                                               &Jaisawal \& Naik (2016b)\\
14 &Cep X-4                 &28, 45                                             &Jaisawal \& Naik (2015b)\\   
15 &4U 0352+309      &29                                              & Coburn et al. (2001)\\  
16 &IGR J16393-4643         &29.3                                            &Bodaghee et al. (2016)\\         
17 &Cen X-3                 &30                                                &Burderi et al. (2000)\\   
18 &IGR J16493-4348         &30                                                   &D'A{\`i} et al. (2011) \\    
19 &RX J0520.5-6932         &31.5                                    &Tendulkar et al. (2014)\\                          
20 &LS V+44 17              &32                                       &Tsygankov et al. (2012)\\
21 &MXB 0656-072            &33                                                   &McBride et al. (2006)\\
22 &XTE J1946+274           &36                                              &Heindl et al. (2001)\\
23 &4U 1626-67              &37                                              &Orlandini et al. (1998)\\
24 &GX 301-2                &37                                                  &Suchy et al. (2012)\\
25 &Her X-1                 &39, 73                                          &Enoto et al. (2008)\\
26 &MAXI J1409-619          &44, 73, 128                                              &Orlandini et al. (2012)\\
27 &1A 0535+262             &45, 100                                           &Terada et al. (2006)\\
28 &GX 304-1                &54                                                &Yamamoto et al. (2011) \\ 
29 &1A 1118-615             &55, 112?                                            &Doroshenko et al. (2010)\\ 
30 &GRO J1008-57            &76                                                   &Yamamoto et al. (2014)\\
\\
\hline
\\

1 &EXO 2030+375            &11? 61?                                         &Wilson et al. (2008)  \\
2 &GS 1843+009             &20?                                             &Mihara et al. (1995)\\
3 &2S 0114+650             &22?, 44?                                       &Bonning et al. (2005) \\
4 &GX 1+4                  &34?                                               &Ferrigno et al. (2007)  \\ 
5 &OAO 1657-415            &36?                                                  &Orlandini et al. (1999)\\
6 &4U~1700-37              &37?                                                   &Jaisawal \& Naik (2015a) \\     
7 &4U~1909+07              &44?                                                  &Jaisawal et al. (2013)\\   
8 &LMC X-4                 &100?                                                   &La Barbera et al. (2001) \\  

\hline
\hline
\end{tabular}
\\
XMMU J05$^*$ stands for XMMU J054134.7-682550 \\

\label{cyc}
\end{table}


\section{Observations and Analysis}

The broadband spectral studies of 4U~1909+07, Cep~X-4 and 
SMC~X-2 were performed by using data from $Suzaku$ (Mitsuda et al. 2007)
 and $NuSTAR$ (Harrison et al. 2013) observatories. 4U~1909+07 and Cep~X-4 
were observed with $Suzaku$ during 2010 November and 2014 July, respectively. 
Observations of SMC~X-2 were carried out with $NuSTAR$ at three epochs during 
2015 October X-ray outburst. The energy spectra of pulsars were described with 
different phenomenological models such as high energy cutoff power-law model (highecut), 
cutoff power-law model, Fermi-Dirac cutoff model (fdcut), negative and positive 
exponential cutoff power-law (NPEX) and more physical Comptonization model such as 
CompTT. The spectral fitting was carried out using XSPEC package. In addition to 
continuum, additional components such as partial covering, Gaussian functions for
emission lines from fluorescence emissions, CYCLABS or Gaussian absorption component 
for cyclotron features were also used whenever required.       
 
\begin{figure}[h]
\centering
\includegraphics[height=7.8cm, width=5.8cm, angle=-90]
{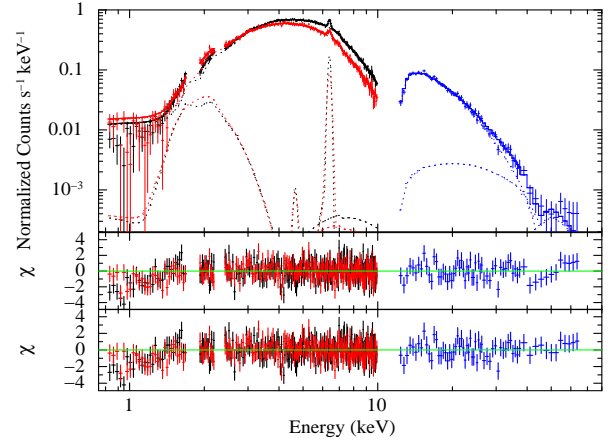}
\caption{The 1-70 keV energy spectrum of 4U~1909+07 fitted with partial
covering NPEX model along with 6.4 keV iron line and a cyclotron line at
44 keV. Middle and bottom panels show the contributions of the residuals 
to $\chi^2$ for each energy bin without and with cyclotron line, respectively.}
\label{fig-1}
\end{figure}

\section{Results}
\subsection{Cyclotron line in 4U~1909+07}

4U~1909+07 is a slow pulsar with spin period of 604 s (Levine et al. 2004). The optical
companion of the X-ray pulsar was found to be an OB supergiant star (Morel \& Grosdidier, 
2005). The orbital period of the system was reported to be 4.4 d. The broad-band X-ray continuum 
of the pulsar was poorly studied in the past due to the dearth of high quality data. We have 
studied the 1-70 keV wide band spectrum of pulsar by using a $Suzaku$ observation. The 
partial covering NPEX and high energy cutoff models were found to describe the continuum well. Apart
from the continuum model, additional components such as a blackbody component for soft X-ray 
excess and a Gaussian function for 6.4 keV iron emission line were also required in the spectral
fitting. While fitting the pulsar spectrum with standard continuum models, an absorption-like 
feature was seen at $\sim$44 keV (see Fig.~\ref{fig-1}). We have investigated the spectrum
with combination of several models to check the possibility of model dependency of the feature. 
However, a broad absorption feature was clearly seen in the pulsar continuum in a model independent 
manner. We identified this feature as the cyclotron absorption line of the pulsar. We attempted to 
normalize the pulsar spectrum with that of Crab pulsar which has a featureless continuum with photon
index of 2.1. The crab ratio clearly showed the presence of absorption feature in 40 to 50 keV range
(see Figure-7 from Jaisawal et al. 2013). 
This also confirmed the presence of cyclotron feature in 4U~1909+07. The statistical tests such as 
F-test and run-test were performed to investigate the significance of the absorption feature. Based 
on these analysis, we reported a probable detection of cyclotron line at $\sim$44 keV in 4U~1909+07. 
Corresponding magnetic field strength of the neutron star was estimated to be 3.8$\times$10$^{12}$~G 
(Jaisawal et al. 2013).

\subsection{Fundamental and first harmonics of cyclotron line in Cep~X-4}

\begin{figure}[t!]
\centering
\includegraphics[height=8cm, width=6.8cm, angle=-90]
{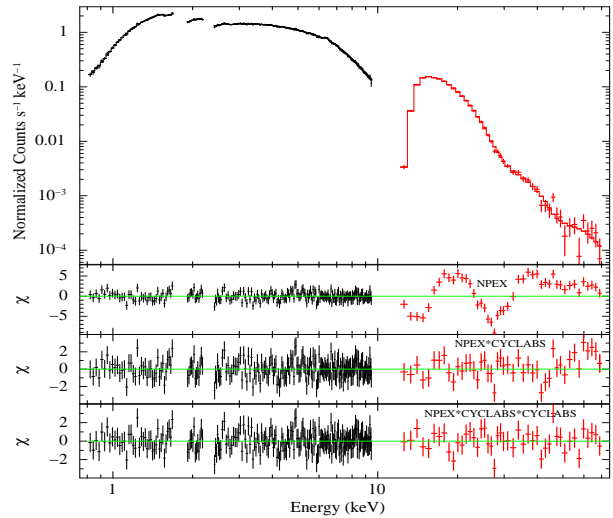}
\caption{The 1-70 keV energy spectrum of Cep~X-4 fitted with partial
covering NPEX model along with 6.4 and 6.9 keV iron lines and cyclotron 
lines. The second, third and fourth panels show the contributions
of the residuals to $\chi^2$ when pulsar continuum was fitted with a partial 
covering NPEX model without, with one and two cyclotron line components,
respectively.}
\label{fig-2}
\end{figure}

Be/X-ray binary pulsar Cep~X-4 was observed with $Suzaku$ during its 2014 June-July Type~I 
X-ray outburst. We have used data from a $\sim$60~ks observation of the pulsar in the declining 
phase of the outburst. Cep~X-4 is a pulsar which have a pulsation period of 66.33 s. A fundamental 
cyclotron line was discovered in the pulsar at $\sim$30 keV with $Ginga$ observation (Mihara et al. 1991). 
In this work, we report the detection of the first harmonic of the cyclotron line. The 1-70 keV spectrum 
obtained from $Suzaku$ observation was well described with standard continuum models such as NPEX, hecut, 
CompTT along with partial covering component and Gaussian functions for two emission lines at 6.4 and 6.9 keV. 
While fitting, a strong absorption feature was detected at 28 keV which has been known as the cyclotron
line in the pulsar. In addition to this, an absorption like feature at $\sim$45 keV was also detected 
(see Fig.~\ref{fig-2}). This feature was seen in the pulsar spectrum while fitting with all the above 
continuum models. We have identified this feature as the first harmonic of the cyclotron line though it 
was detected at 1.7 times of fundamental line energy. The statistical significance of the absorption line
was also tested by using XSPEC script `simftest'. We found the detection of feature with statistical 
significance of $>$4$\sigma$.

Generally, harmonics of the cyclotron line is expected at the multiples of
fundamental energy. However, in case of Cep~X-4, we have detected the 
first harmonic below the ideal coupling factor 2. Anharmonicity in line 
energy has also been seen in other pulsars, where line energies ratio was 
found at above as well as below 2 (see Table~\ref{cyc} and references therein). 
Relativistic approximation of photon-electron scattering can contribute 
to the anharmonicity in cyclotron line energies (M{\'e}sz{\'a}ros 1992). However, such low value of
line ratio, as seen in Cep~X-4, is difficult to explain in this approximation. There 
are other possibilities by which anharmonicity in cyclotron line energies can be 
understood. It is probable that the fundamental and harmonic lines are formed at two 
different scale heights which can have different optical depth. As a result, anharmonicity 
in the line energy ratio can be observed. Studies on cyclotron lines from Nishimura (2005) 
showed that increased in magnetic field in line forming region can produce the 
anharmonic cyclotron line energies. The effect of viewing column/line forming region 
at large angles can also be the cause (Nishimura 2013). Alternatively, anharmonicity 
in line energies may indicate the distortion or changes in the magnetic field geometry. 
However, we have not detected any significant variation in cyclotron line parameters 
with the pulsar-phases in phase-resolved spectroscopy (see Figure-4 from Jaisawal \& 
Naik, 2015b).

\subsection{Discovery of cyclotron line in SMC~X-2}

\begin{figure}[htb!]
\centering
\includegraphics[width=7.cm, angle=-90]
{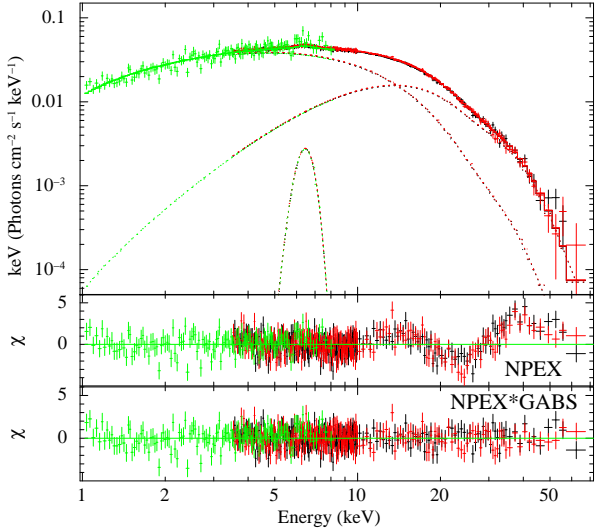}
\caption{The 1-79 keV energy spectrum of SMC~X-2 obtained from first 
$NuSTAR$ and $Swift$/XRT at the peak of the outburst, fitted with partial covering NPEX 
model along with 6.4 keV iron line and cyclotron line at 27 keV.
The second and third panels show the contributions of the residuals 
to $\chi^2$ when pulsar continuum was fitted with a partial covering 
NPEX model without and with cyclotron line component, respectively.}
\label{fig-3}
\end{figure}

SMC~X-2 is a high mass X-ray binary pulsar with a spin period of 2.37 s 
(Corbet et al. 2001). The orbital period of the binary system was estimated 
to be 18.4 d (Townsend et al. 2011). This source went in to an intense
outburst in 2015 September and observed with major X-ray observatories. We have
investigated spectral properties of the pulsar with $NuSTAR$ and $Swift$/XRT at three 
different luminosity epochs ($>$10$^{38}$~erg/s). The 1-79 keV spectra were well 
fitted with the traditional pulsar continuum models such as NPEX, fdcut, cutoff 
power-law along with iron emission line. While fitting, an additional absorption 
like feature was also detected in the spectra at $\sim$27 keV for the first time
(see Fig.~\ref{fig-3}). This feature was seen in a model independent manner 
and identified as a cyclotron absorption line in the pulsar spectrum. Corresponding 
to the line energy, the magnetic field of the neutron star was estimated to be 
2.3$\times$10$^{12}$~G (Jaisawal \& Naik, 2016b). 

$NuSTAR$ observations also provided opportunity to probe the luminosity dependence
of the cyclotron line in this pulsar. As the observations were performed at 
three different luminosity levels, we detected a marginal changes in the cyclotron line
energy that was anti-correlated with the luminosity (Table~2 in Jaisawal \& Naik, 2016b).  
These luminosity dependent variation in cyclotron line energy can be attributed to
the changes in the line forming region. There are a couple of sources which 
show a positive correlation of cyclotron line energy with luminosity (e.g. GX 304-1; 
see discussion section in Jaisawal et al. 2016a). It is believed that these correlations 
are attributed to a critical luminosity. The pulsars below the critical luminosity show 
a positive correlation and vice-versa. In case of SMC~X-2, the source luminosity 
($>$10$^{38}$~erg/s) was in the super-critical regime where the radiation pressure 
dominates and a shock can be formed above the neutron star surface (Becker et al. 2012). 
As the luminosity increases, the line-forming region is expected to shift in the accretion 
column. This results a negative correlation between cyclotron line energy and luminosity.

\section{Summary}
 
We have reported the detection of cyclotron line(s) in pulsars 
such as 4U~1909+07, Cep~X-4 and SMC~X-2 by using high quality data 
from $Suzaku$ and $NuSTAR$ observatories. We have also probed  
various phenomena such as anharmonicity in line energies, luminosity 
dependent characteristics as well as mapped the magnetic field around pulsars.
The presence of anharmonic lines in Cep~X-4 may indicate that the fundamental 
and first harmonic of cyclotron line were not forming at the same scale. It is 
also possible that the line-forming regions are viewed at larger angles. For the
first time, we detected a cyclotron line in SMC~X-2 along with a negative correlation
between cyclotron line energy and luminosity and discussed these findings in term of 
changes in the line-forming region.

\label{last}

\end{document}